# Trajectory Retrieval and Component Investigations of Southern Polar Stratosphere Based on High Resolution Spectroscopy of Totally Eclipsed Moon Surface

Oleg S. Ugolnikov[1], Anna F. Punanova[2], Vadim V. Krushinsky[2]

[1]Space Research Institute, Russian Academy of Sciences,
Profsoyuznaya st., 84/32, Moscow, 117997, Russia
[2]Kourovka Astronomical Observatory, Ural Federal University,
Lenina st., 51, Ekaterinburg, 620000, Russia

E-mail: ougolnikov@gmail.com, punanovaanna@gmail.com, krussh@gmail.com.

**Abstract.**

In this paper we present the high resolution spectral observations of the fragment of lunar surface during the total lunar eclipse of December 10, 2011. The observations were carried out with the fiber-fed echelle spectrograph at 1.2-m telescope in Kourovka Astronomical observatory (Ural mountains, central Russia). The observed radiation is transferred by tangent trajectory through the southern polar stratosphere before the reflection from the Moon and spectra contain a number of absorption bands of atmospheric gases ($O_2$, $O_3$, $O_4$, $NO_2$, $H_2O$). High resolution analysis of three $O_2$ bands and $O_4$ absorption effects is used to trace the effective trajectory of solar emission through the stratosphere and to detect the contribution of scattered light. Bands of other gases allow us to measure their abundances along the trajectory.

**1. Introduction.**

A lunar eclipse is an astronomical event characterized by unique geometry of the radiation transfer [1]. Direct solar emission can not reach the surface of the Moon immersed into the umbra (shadow) of the Earth. But it can be refracted in the Earth's atmosphere and enter the geometrical shadow area. That is why the Moon does not fade in the sky during the total eclipses. Along with the refraction, light scattering and absorption take place in the atmosphere. Since these processes can depend on the wavelength, the color of the eclipsed Moon changes and strong atmospheric lines appear in the lunar spectrum. The one is similar to the spectrum of a distant star of Sun-like spectral class during the transit of the planet with dense and optically thick atmosphere [2, 3]. In the case of the lunar eclipse these lines are sensitive to the component concentration near the ray perigee altitude. Normally, tangent transmission spectrum of the Earth's atmosphere can be observed from space, which became the basis of satellite technique of atmosphere composition measurements [4, 5]. Lunar eclipses provide the unique opportunity to carry out such measurements from the ground.

The basis of radiation transfer theory during the lunar eclipses is established in [1]. Refracted solar emission makes the principal contribution to the brightness of the lunar surface. This simplifies the theory and makes it possible to retrieve the additional aerosol and trace gases extinction in the different layers above the limb [6-8]. However, the emission scattered in the atmosphere can be also noticeable. Theoretical estimation of its contribution is performed in [9]. It is found to be sufficient for the wavelengths below 450 nm or in the volcanically perturbed atmosphere, which is not a rare case: the signs of scattered light were found during the eclipse of August, 16, 2008 [10]. Authors relate these signs to Kasatochi volcano eruption in Alaska (quite close to the limb) occurred just before the eclipse.

Photometric analysis of a number of eclipses [6-8] in red and near-infrared spectral range out of atmospheric gases absorption bands had shown that the brightness of the outer umbra part (less than 0.2° from the umbra edge) is usually close to the theoretical value for the gaseous atmosphere model without the aerosol extinction, revealing the clear atmosphere conditions above 10 km. The exceptions occur in the equatorial and tropical atmosphere. The most remarkable one is the eastern tropical umbra part illuminated through the South-Asian troposphere during the eclipse of June, 15, 2011, the darkest through the last years [11].

Results of spectral measurements of lunar eclipses are interesting since they are sensitive to the variations of atmosphere components along the tangent path above the limb. This path is long (about several hundreds kilometers), and its sufficient and optically thick fraction lies in the horizontal layer with almost constant altitude. It provides high accuracy and good vertical resolution of limb atmosphere spectroscopy. It is especially true for stratosphere components, first of all, stratospheric ozone. Chappuis lines of ozone define the general brightness and color characteristics of the eclipsed Moon although they seem to be quite weak in a transmission spectrum in a vertical atmosphere column.

The spectrum of the lunar surface was obtained with high resolution during the partial eclipse of August, 16, 2008 [3], but in the penumbra, where the Moon is partially illuminated by direct solar radiation transferred above the atmosphere. Umbral area was measured once during the same eclipse in a wide spectral range [2], but with lower spectral resolution (about 1000). In this paper we analyze a number of high resolution spectra of the lunar surface in umbra obtained during the total eclipse of December, 10, 2011. This time the moonlight did not strongly increase the flux from the surrounding sky background. This eclipse is especially interesting since the solar radiation was transferred through the Antarctic stratosphere not far from seasonal ozone depression. Long duration and less depth of the eclipse allowed a thorough investigation of the Antarctic stratosphere.

The rest of the paper is organized as follows. Section 2 contains the description of observations. In Section 3 a low-resolution analysis is performed, the atmospheric species with broad absorption bands ($O_3$, $O_4$, and $NO_2$) are studied. High-resolution analysis of $O_2$ lines and trajectory retrieval is presented in Section 4. Investigations of water vapor are discussed in Section 5. Finally, Section 6 contains a conclusion.

**2. Observations.**

Spectral observations of the total lunar eclipse of December, 10, 2011 were conducted in Kourovka Astronomical Observatory, Russia (57.0°N, 59.5°E). The observations were carried out with the optical fiber-fed echelle spectrograph at the Nasmyth focus of 1.2-m telescope. The spectral resolution is about 30000. Fiber diameter corresponds to the angular size 5″. Instrumental band covers the wavelength range from 410 to 780 nm.

The calibration frames (bias, flat field, ThAr spectrum) were collected once per hour. We used the ThAr lamp spectrum for wavelength scale calibration. The accuracy of this procedure is 0.0003 nm. The narrow lines of the ThAr lamp were used to build the instrument point-spread function (PSF) of the spectrograph. We did not obtain the spectra of the sky because it is not necessary for high-resolution spectra. The sky emission lines are weak during the totality and their flux is negligible (the exposure must be about ten times more than the totality duration to observe the sky lines with a high-resolution spectrograph). All frames were processed with IRAF/echelle package [12]. Ambient temperature and humidity were recorded at about five-minute intervals to control the weather conditions stability during the observations.

| Spectrum № | UT (middle) | | Zenith distance | Distance to the umbra edge |
|---|---|---|---|---|
| | hr | min | ° | ° |
| 1 | 14 | 14.2 | 68.2 | 0.086 |
| 2 | 14 | 20.2 | 67.4 | 0.095 |
| 3 | 14 | 26.3 | 66.7 | 0.101 |
| 4 | 14 | 32.2 | 65.9 | 0.102 |
| 5 | 14 | 38.1 | 65.1 | 0.100 |
| 6 | 14 | 44.4 | 64.3 | 0.093 |
| 7 | 14 | 50.2 | 63.5 | 0.083 |
| 8 | 14 | 56.1 | 62.7 | 0.069 |

*Table 1. Parameters of the observed spectra of the lunar eclipse.*

We acquired the spectra of the region of the lunar surface with selenographic coordinates 50°S, 5°W, south-eastwards from bright crater Tycho. This place is characterized by almost uniform albedo. During the eclipse the place was traversing the southern edge of umbra. The radiation transferred through the lower Antarctic stratosphere contributed the major portion of brightness of the observed spot. Eight frames (exposure is 300 seconds for the each one) were obtained as this surface moved through the umbra. The frames were not combined to improve the signal-to-noise ratio (S/N). The spectra with eclipsed Moon have the S/N ratio up to 65 in different orders, which is sufficient for this study. The S/N ratio near H$\alpha$ line (656.3 nm) is about 45. The observational parameters of recorded spectra are presented in Table 1. The positions of the observed spot in the umbra during the exposures are shown in the Figure 1. The limb point was moving eastwards along the Antarctic shore. The mid-exposure positions of the limb point are depicted in the Figure 2.

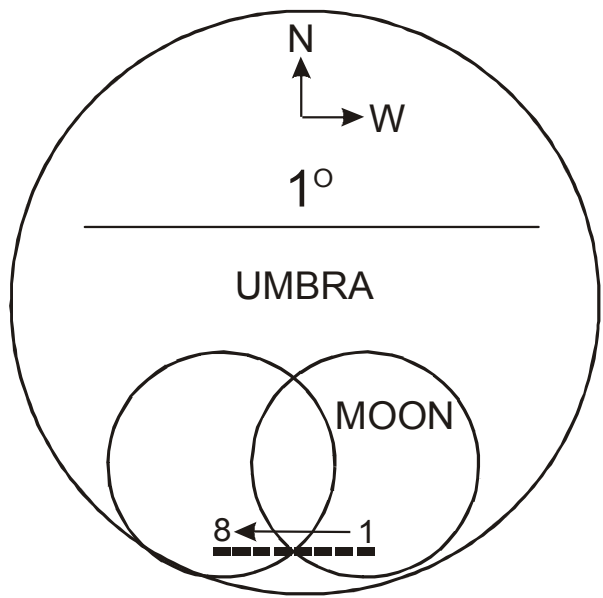

*Figure 1. Positions of the observed point of the lunar surface inside the umbra in the sky (west direction is to the right). The Moon position is shown for the first and the last eclipse frames. The geometrical umbra border (the atmosphere-free case) is not expanded by the factor 1.02, what is usually done in astronomical ephemeris.*

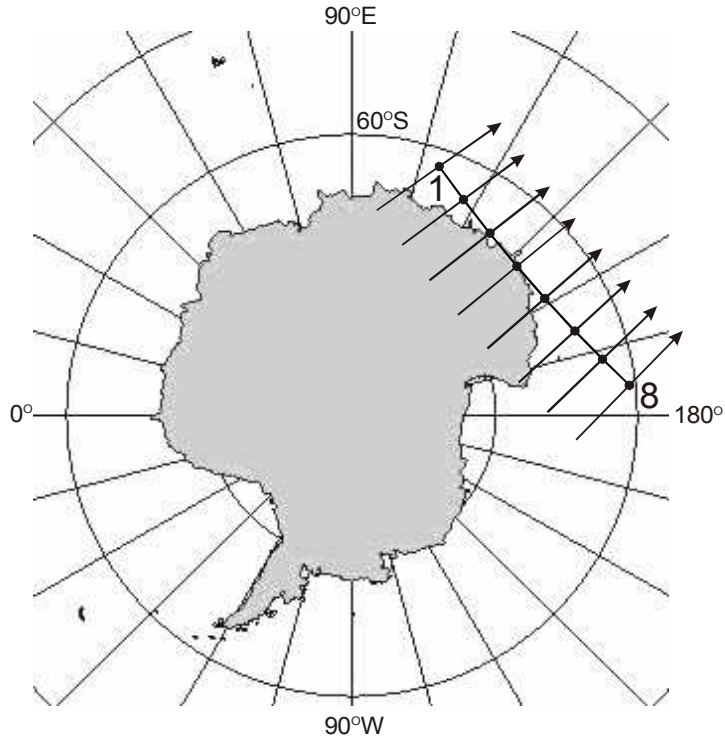

*Figure 2. The motion of the limb point below the light path perigee during the observations. Arrows show the direction of tangent emission propagation.*

To take the lunar albedo spectral dependency into account we obtained the spectra of the same spot on the lunar surface after the end of the eclipse. The Moon was ascending during the observations. Throughout the eclipse zenith angle decreased from 70° to 60° and after the end of the eclipse it was about 40°. The transparency of the atmosphere over the observatory depends on the wavelength too. To take it into account, the spectra of a standard star were obtained between the totality and full Moon observations. The standard star 109 Tauri was close to the Moon during the observational period prior to its lunar occultation. The star has almost the same spectral class as the Sun, but spectral lines of the star are shifted a little due to the star radial velocity (about +18 km/s).

**3. Spectrum of the tangent limb transmission. $O_3$, $O_4$, and $NO_2$ absorption.**

Far from the range of fine structure atmospheric absorption (first of all, $O_2$ and $H_2O$ bands) the observed spectrum of the eclipsed Moon can be described by the equation:

$$F_E(\lambda, z) = S(\lambda) E(\lambda) L(\lambda) A(\lambda)^{M(z)} \qquad (1),$$

where $S$ is the solar spectrum, $E$ is the atmosphere transmission spectrum along the tangent trajectory above the limb, $L$ is the lunar albedo, $A$ is the local vertical atmosphere transmission spectrum, and $M$ is the atmospheric mass. For zenith angles $z<70°$ the last value is equal to $1/\cos z$. Analogous expression for value $F_0(\lambda, z)$ outside the umbra and penumbra has the following form:

$$F_0(\lambda, z) = S(\lambda) L(\lambda) A(\lambda)^{M(z)} \qquad (2).$$

The spectrum of standard star is

$$F_S(\lambda, z) = C\, S_1(\lambda) A(\lambda)^{M(z)} \qquad (3),$$

where $S_1$ is the natural spectrum of the star and $C$ is the constant defined by the size of star image and its position in the focal plane relative to the fiber. This value varies for different star spectra. If the eclipsed Moon spectrum is obtained at zenith angle $z_E$ and the full Moon spectrum is obtained at zenith angle $z_0$, then we have:

$$E(\lambda) = \frac{F_E(\lambda, z_E)}{S(\lambda) L(\lambda) A(\lambda)^{M(z_E)}} = \frac{F_E(\lambda, z_E)}{F_0(\lambda, z_0)} A(\lambda)^{M(z_0) - M(z_E)} \qquad (4).$$

Having two spectra of the standard star at the zenith angles $z_1$ and $z_2$, we express their ratio:

$$\frac{F_{S1}(\lambda, z_1)}{F_{S2}(\lambda, z_2)} = \frac{C_1}{C_2} A(\lambda)^{M(z_1) - M(z_2)} \qquad (5).$$

Substituting (5) into (4), we have:

$$E(\lambda) = C_{12} \frac{F_E(\lambda, z_E)}{F_0(\lambda, z_0)} \cdot \left( \frac{F_{S1}(\lambda, z_1)}{F_{S2}(\lambda, z_2)} \right)^{\frac{M(z_0) - M(z_E)}{M(z_1) - M(z_2)}} \qquad (6).$$

Here $C_{12}$ is the constant derived from the ratio ($C_1/C_2$). It is also affected by the slant change of lunar albedo due to the difference of phase angles inside and outside the umbra [13]. Due to these factors, there is no absolute radiometric calibration, and the spectrum $E(\lambda)$ can only be found in arbitrary units.

Figure 3 shows the spectrum $E(\lambda)$ for the middle of totality (position 4 in the Figure 1), all other obtained spectra have the same structure. The general properties are analogous to the spectra of August, 2008 eclipse [2, 10]. Strong solar spectrum lines (as Hα and Na doublet) are the same for the eclipse and non-eclipse spectra of the Moon and disappear in the transparency spectrum. The function $E(\lambda)$ increases with the wavelength for $\lambda > 600$ nm, defining the red color of the eclipsed Moon. The variations including two minima near 600 nm are basically related with spectral profile of ozone Chappuis absorption bands. $NO_2$ bands and effects of $O_4$ absorption around 477 and 577 nm are also noticeable.

The $O_4$ molecule is the merging product of two $O_2$ molecules and its concentration is known to be proportional to the $[O_2]$ squared. The absolute values of $O_4$ cross-section depend on the unknown constants of the merging and dividing reactions, but the absorption spectrum per squared $O_2$ concentration (units of $cm^5$/molecule$^2$) is known [14]. Owing to its well-known spatial distribution, $O_4$ is a good tool for the trajectory retrieval and estimation of aerosol and multiple scattering [15]. Since these processes depend on the wavelength, the $O_4$ column densities can vary for different spectral bands. This effect was already noticed in the spectra of the lunar eclipse [10], however, one of two bands used there (630 nm) is seen as background behind $O_2$ lines system and thus hard to study numerically. The $O_2$ bands themselves are better to analyze by high resolution spectroscopy. The results of the analysis are presented below.

To fix the bands of $O_3$, $O_4$, and $NO_2$, we fit the spectrum by the following model using the least squares method:

$$\ln E(\lambda) = \sum_{n=0}^{3} K_n \lambda^n - Q_{O3} s_{O3}(\lambda) - Q_{NO2} s_{NO2}(\lambda) - P_{O41} n_{O41}(\lambda) - P_{O42} n_{O42}(\lambda) \qquad (7).$$

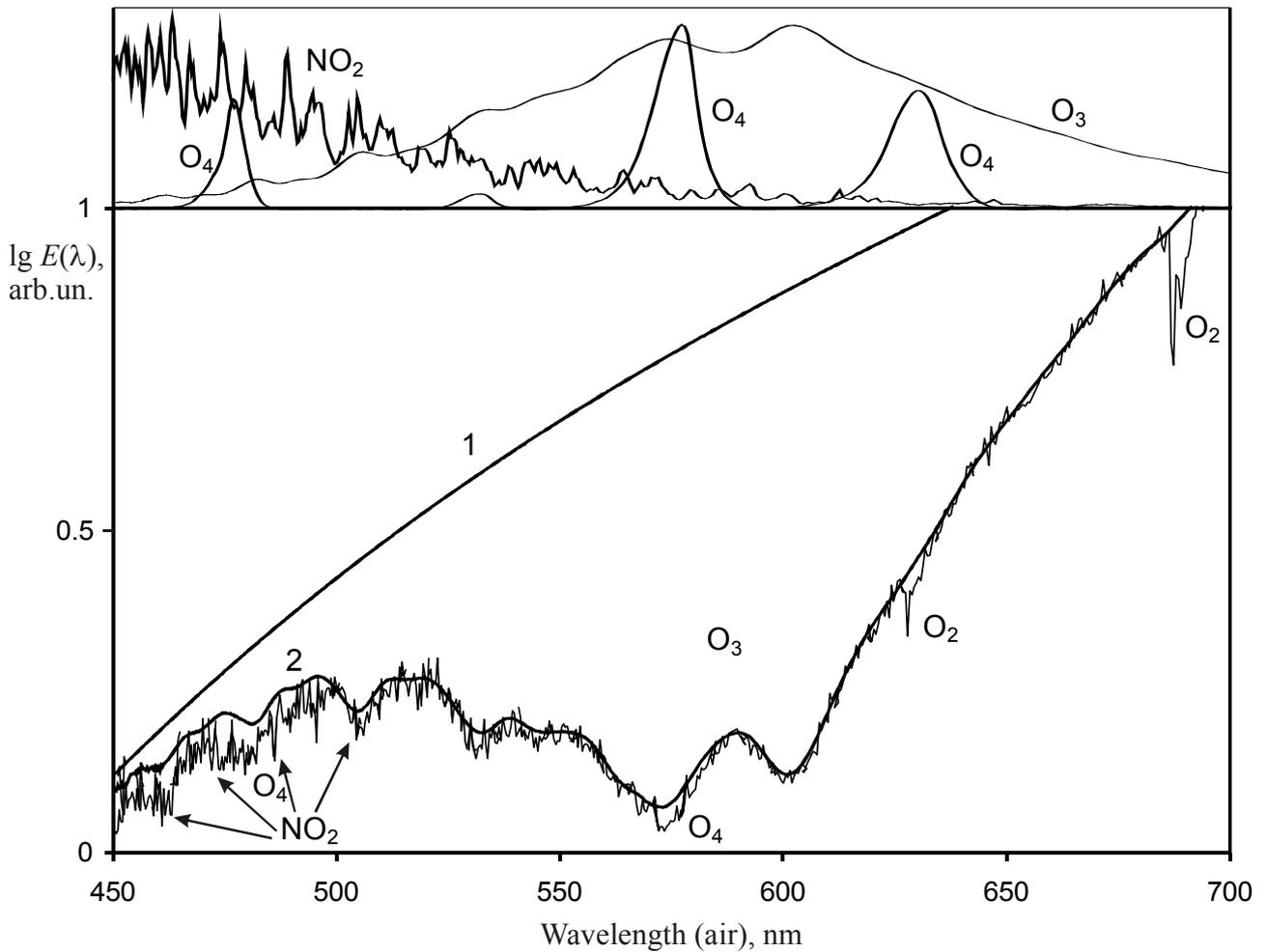

*Figure 3. The transmission spectrum of polar stratosphere in the middle of the eclipse (spectrum 4). The approximation curve 1 corresponds to the non-absorbing gaseous atmosphere, the curve 2 includes the ozone absorption. Upper panel shows the wavelength dependencies of $O_3$, $O_4$, and $NO_2$ cross-sections.*

Here $s_{O3}(\lambda)$ is the cross-section of ozone molecule and $Q_{O3}$ is the number of $O_3$ molecules per square unit along the emission trajectory. Analogous denotations are made for $NO_2$. The spectrum of $O_4$ absorption coefficient per squared $O_2$ concentration $n_{O4}(\lambda)$ is separated in two parts: the first one is non-zero for shorter wavelengths $\lambda<500$ nm, while the second one is for $\lambda>500$ nm. The boundary wavelength value (500 nm) is defined by total vanishing of $O_4$ absorption. The quantities $P_{O41}$ and $P_{O42}$ are integrals of $[O_2]$ squared along the path defined by 477 and 577 nm absorption bands of $O_4$, respectively.

The spectral range of $O_2$ and $H_2O$ absorption bands was not included to this procedure, so the $O_4$ absorption band at 630 nm (obscured by $O_2$ lines) was not actually taken into account. The values of cross-sections of $O_3$ and $NO_2$ for polar stratosphere conditions are taken from [16] and [17], respectively. Result of the procedure is also shown in the Figure 3. Ozone absorption profile fits the observed spectral features of $E(\lambda)$ except the $O_2$ bands in the right, $O_4$ and $NO_2$ bands add just a little correction visible in the figure. The resulting value of $P_{O41}$ (about $10^{44}$ molecule$^2$/cm$^5$ in the middle of totality) is surely about 2 times higher than $P_{O42}$ ($5\cdot10^{43}$ molecule$^2$/cm$^5$), showing the different radiation transfer geometry at 477 and 577 nm and the scattered light contribution in the first of these bands. This question will be addressed below (Section 4).

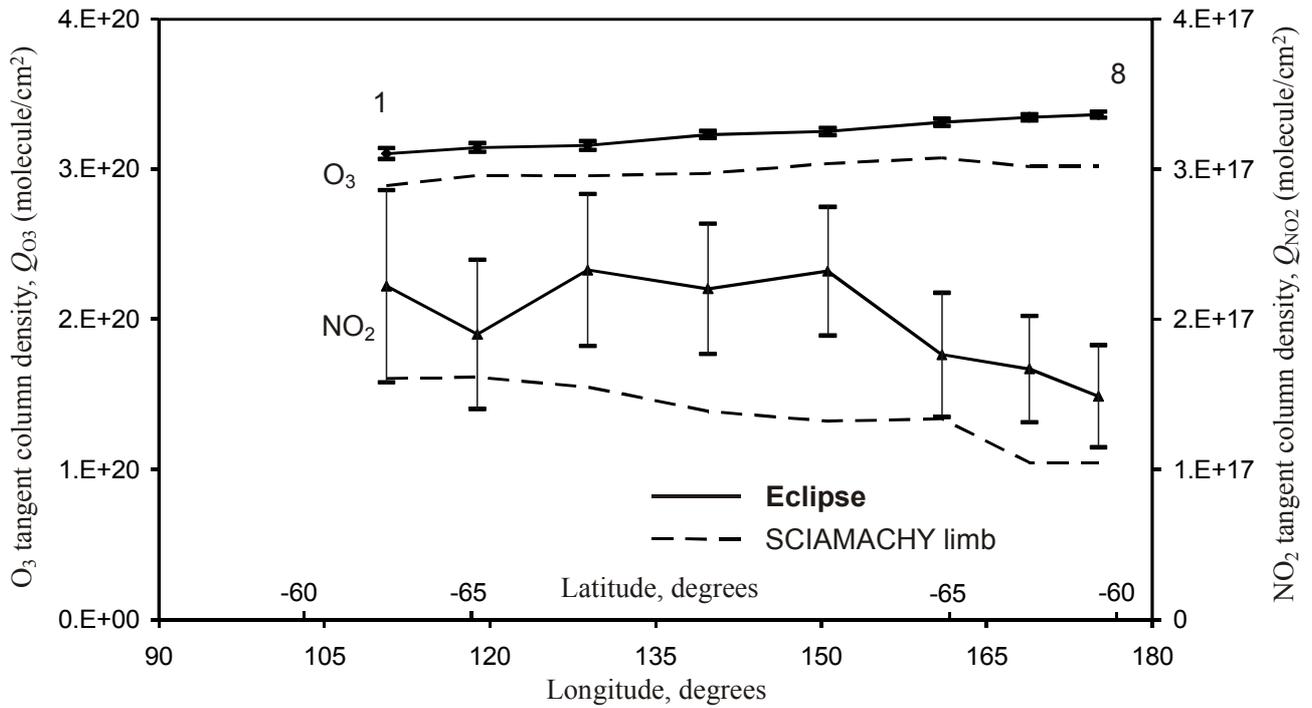

*Figure 4. The tangent column densities of $O_3$ and $NO_2$ along the effective path for different point positions inside the umbra compared with SCIAMACHY limb data.*

The ozone Chappuis absorption is found to be strong. Comparing the curves with and without ozone (the same figure), we see that the lunar surface brightness at 600 nm is reduced by a factor of 6. The dependences of $O_3$ and $NO_2$ tangent column densities on the coordinates of the Antarctic limb points are shown in the Figure 4. The changes of $NO_2$ abundance seems to be related to its diurnal variations. The first spectra correspond to the evening part of the limb (Sun is setting), the last spectra – to the morning one (Sun is rising). The number of $NO_2$ molecules is naturally anti-correlates with the number of $O_3$ molecules. The accuracy of the estimation of $NO_2$ abundance is not so good. The reasons for that are the low intensity of eclipsed Moon in the blue part of the spectrum where $NO_2$ absorption takes place, and influence of scattered light for $\lambda < 500$ nm.

The relative accuracy of the ozone abundance determination is much better. We see that ozone concentration in the lower stratosphere is slowly increasing eastwards. It partially agrees with the behavior of total ozone column density by SCIAMACHY TOSOMI data [18]. To compare the trajectory results with available satellite data on the same day, the retrieval of trajectory parameters is needed, which will be done below.

**4. Effective trajectory retrieval.**

Figure 5 shows the principal scheme of radiation transfer during the lunar eclipse. For each moment, the position of the Earth's limb can be calculated. Given the position angle of the Moon surface element in the umbra, we find the coordinates of the limb point lying in the plane including the Moon surface point being measured and the centers of the Earth and the Sun. It is the perigee projection on the surface of the Earth.

Solar emission crosses the atmosphere twice (paths 1 and 2 in the figure). The object of interest is the fraction of tangent trajectory path through the southern polar atmosphere (path 1). It is characterized by the perigee altitude $h_0$ or the corresponding pressure $p_0$. Given the model of vertical temperature distribution, we can calculate the refraction angle $\rho_0$ for this path.

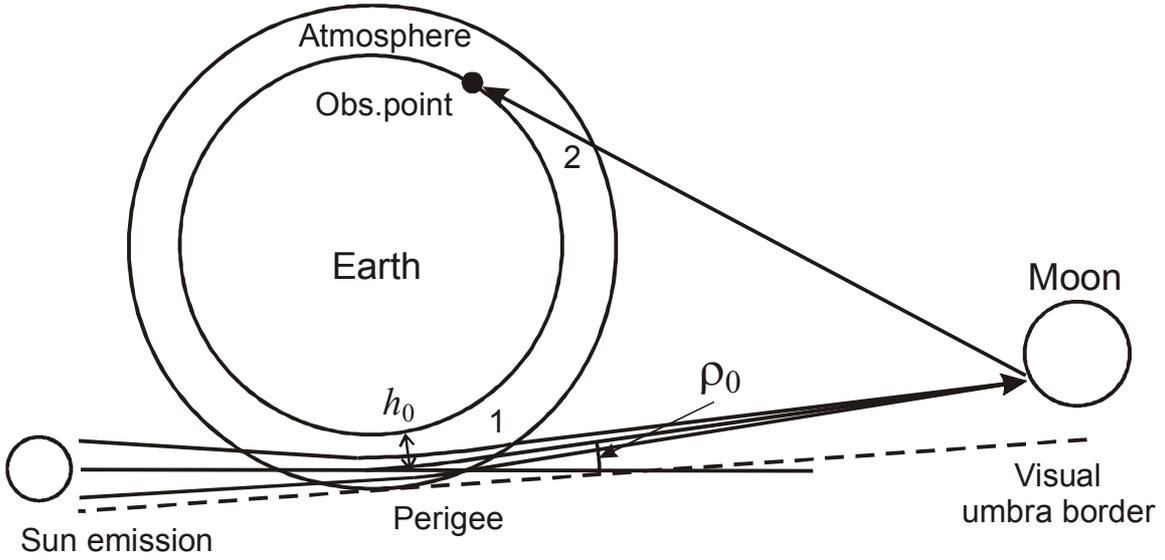

*Figure 5. The scheme of radiation transfer during the lunar eclipse.*

The lunar surface observation point was about 0.1° inside the umbra edge. Adding the solar angular radius, we can expect that the effective refraction angle is not more than 0.3-0.4° (the value can be also decreased by the aerosol extinction in the lower atmosphere), showing that the effective trajectory perigee lies in the lower stratosphere. The temperature does not sufficiently vary with altitude there. The characteristic temperature for this season and location (220 K) is based on the MIMOSA data [19]. The atmosphere absorption takes place mostly along the lowest trajectory fraction near the perigee with the altitudes just a little bit more than the perigee altitude $h_0$. Characterizing the path by the perigee pressure $p_0$, we can use the isothermal model of lower stratosphere. To calculate the transmission spectrum, we assume the path to have the constant pressure $p_0$ and the length giving the same number of molecules along the path. This length can be calculated by the formula:

$$D_0 = B\sqrt{2\pi R H} \qquad (8).$$

Here $R$ is the Earth's radius, $H$ is the atmospheric scale height for the temperature 220 K. The factor $B$ slightly exceeds the unity and characterized by the curvature of the light trajectory. It depends on the perigee pressure, being close to 1.1 for the lower stratosphere.

Effective trajectory of radiation transfer can be retrieved by the spectral lines of atmospheric component with well-known spatial distribution, such as molecular oxygen $O_2$. Instrumental band includes three $O_2$ absorption bands with the wavelengths 630, 690, and 765 nm. Taking the parameters of each line from HITRAN database [20] and building the line profiles using the SPECTRA system [21] with account PSF of our detector $Q(\lambda,\lambda_i)$, we can calculate the spectrum of eclipsed Moon in each of three bands:

$$F_{EL}(\lambda) = \int_{-\lambda_{PSF}}^{\lambda_{PSF}} S(\lambda + \lambda_i) E_L(p_0, D_0, \lambda + \lambda_D + \lambda_i) L(\lambda + \lambda_i) A_L(z, \lambda + \lambda_i) Q(\lambda, \lambda_i) d\lambda_i$$

(9).

The denotations are analogous to the equation (1), $E_L$ and $A_L$ are the atmospheric transmission spectra in $O_2$ bands by trajectories 1 and 2, respectively. Due to the fine structure of the lines at the scale comparable or narrower than the instrumental profile, the eclipsed Moon spectrum in $O_2$ bands cannot be considered as the simple multiplication of transmission spectra by the paths 1 and 2 (as it was done above for $O_3$, $O_4$, and $NO_2$) and must be built by convolution of both transmission spectra and instrumental band $Q(\lambda,\lambda_i)$. We also take into account the tiny Doppler shift of transmission spectra ($\lambda_D$). This shift is due to the double radial velocity of the Moon and motion of the observer towards the Moon together with rotating surface of the Earth. Combined effect has the value about 0.4 km/s. It is less than instrumental resolution but comparable with fine structure scale of $O_2$ spectrum and, thus, slightly changes the final result.

The $O_2$ transmission spectrum of atmosphere layer above the observer $A_L$ is defined by the zenith angle of the Moon $z$ and total amount of $O_2$ in the vertical atmosphere column, and slightly depends on the distribution of the troposphere temperature. Since the relative abundance of $O_2$ does not vary with altitude in the troposphere, the spectrum can be easily integrated by the atmosphere layers with equal step by pressure value.

Formula (9) is used to run the best-fit procedure to derive the effective perigee pressure $p_0$ of the path through the polar stratosphere. To do this, we take the solar spectrum $S(\lambda)$ from [22] and consider the spectral dependencies of lunar albedo $L(\lambda)$, atmosphere absorption (including $O_3$ and $O_4$ bands) and scattering to be polynomial (second degree) within each of three narrow ranges of $O_2$ absorption.

Another way to retrieve the effective radiation path through the polar atmosphere is $O_4$ absorption discussed in the previous chapter which gives the integral values of $[O_2]$ squared along the path. This procedure adds two more spectral intervals (477 and 577 nm) expanding the analysis to the most part of observed spectral range. The dependence of the pressure $p_0$ on $O_4$ abundance is quite weak due to the rapid decrease of the latter value with the altitude, so we can estimate $p_0$ basing even on a poorly-defined $O_4$ absorption.

Figure 6 displays the eclipsed Moon spectrum $O_2$ bands in the range from 759 to 766 nm compared with the best-fit curve and the one for uneclipsed case at the same zenith distance. The results of the best-fit procedure for three systems of $O_2$ lines are shown in the Figure 7. The typical values of the effective pressure for these bands are about 0.16-0.20 atm. The corresponding altitude (11-12 km) is above the Antarctic tropopause level. The results of $O_4$ analysis are shown in the same figure.

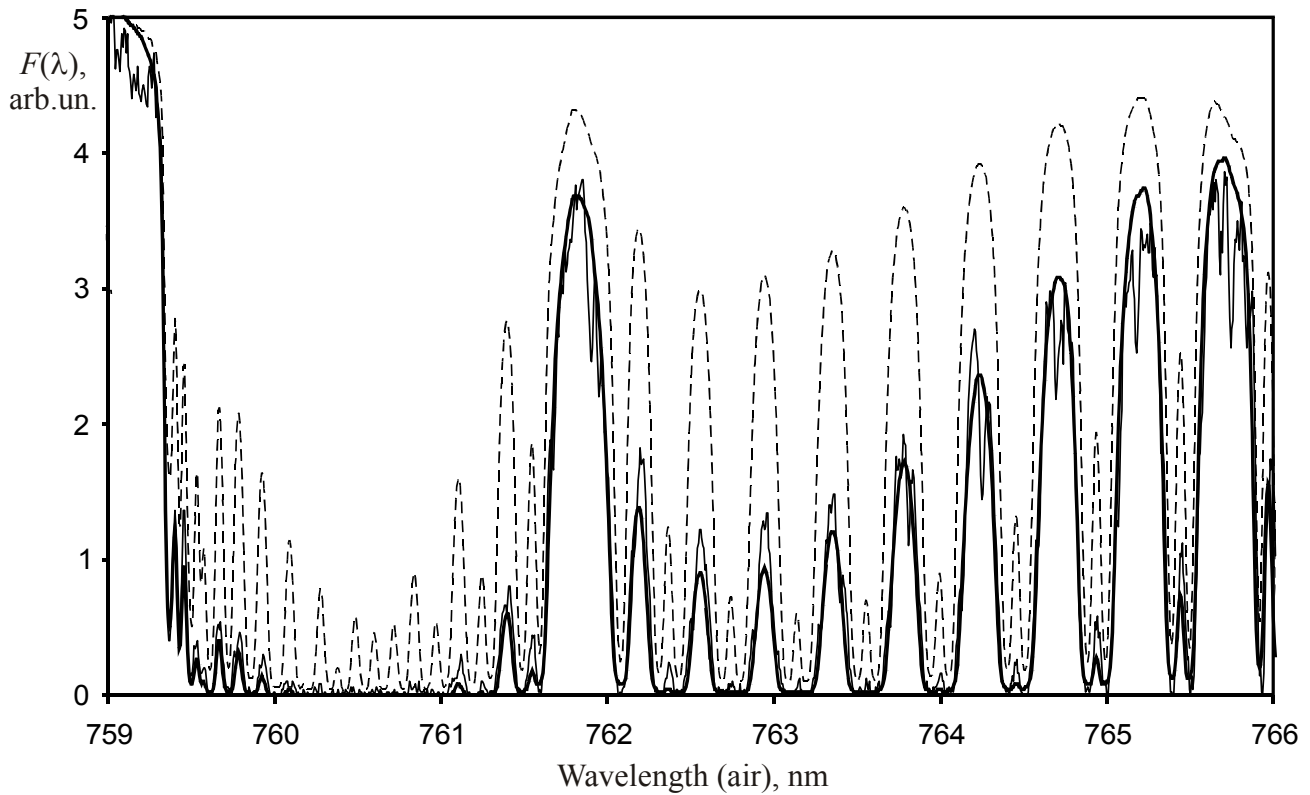

*Figure 6. The fragment of spectrum of the lunar surface 4 with $O_2$ bands near 765 nm. The solid line is the best-fit model, the dashed one corresponds to the local atmosphere absorption only.*

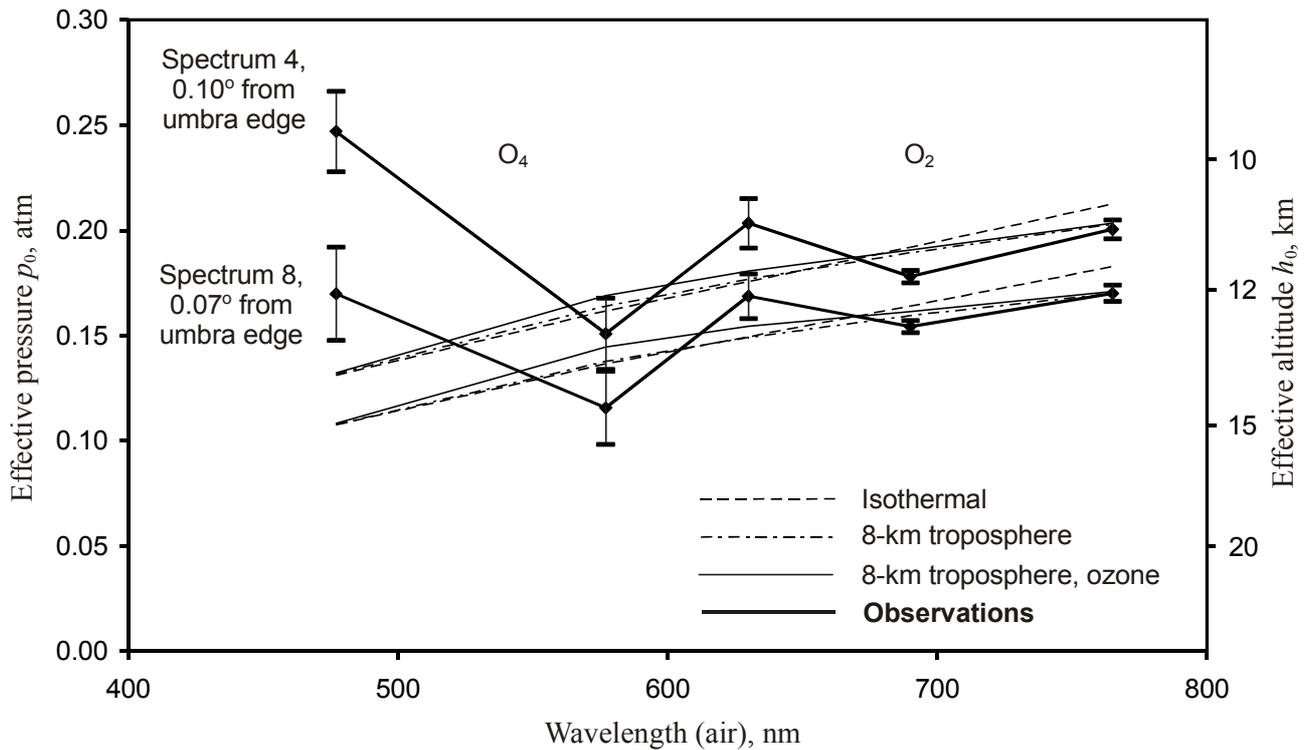

*Figure 7. Wavelength dependency of effective ray perigee pressure and altitude (scale for 8-km troposphere model) by $O_2$ and $O_4$ bands observations (spectra 4 and 8) compared with different atmosphere aerosol-free models.*

Figure 7 presents the results of spectra 4 and 8, obtained at maximal and minimal angular distances from the umbra border, respectively. The behavior of $p_0$ values for other spectra is quite regular: the value increases as the point immerses into umbra and decreases after the middle of totality. It was expected since the refraction angle increases with the perigee pressure.

The figure also contains theoretical dependencies of the same value. First of these dependencies is calculated for the isothermal ozone-free atmosphere with temperature 220 K. Second model also contains 8-km troposphere with linear temperature gradient 5°/km and ground temperature 260 K (the shift of the stratosphere altitude-pressure dependencies between these models is about 0.7 km). The third theoretical curve corresponds to the model with ozone Chappuis absorption for the typical ozone vertical distribution for this date and location based on SCIAMACHY limb measurements [5, 23].

One can see that all model results are quite close to each other, showing small influence of troposphere temperature variations and even ozone absorption on the $p_0$ value. Theoretical curves are also close to the observational ones for all wavelengths except 477 nm. This suggests the lack of aerosol extinction in the whole range of altitudes, where the radiation of different Sun points is refracted (8 km and above). This extinction would decrease the contribution of deeply immersed parts of the Sun with strong absorption lines and, thus, weaken these lines in the observed spectrum and decrease the $p_0$ value. However, this effect was not observed.

In the 477 nm $O_4$ band the reverse effect takes place – the effective pressure $p_0$ exceeds the gaseous value, especially deep inside the umbra (upper curve in the Figure 7). The effect was found above as the increased $P_{O41}$ value. As it was shown theoretically [9], the scattered (Rayleigh and aerosol) light contribution becomes noticeable at such wavelengths (or even at longer ones if the atmosphere is volcanically perturbed, but it is not the case of December 2011 eclipse). The trajectory of scattered light varies depending on the location of scattering point. $O_4$ analysis shows that the optical path of scattered light is longer than the one for refracted light, and scattering trajectory lies below or (most probably) aside the refracted trajectory (however, theoretical analysis [9] had shown that the effective scattering takes place higher). The path can be longer or reach lower atmospheric layers with increased concentration of $O_2$ and, especially, $O_4$. Composition of both effects leads to the sufficient shift of $p_0$ in the Figure 7.

The typical value of $p_0$ for the range of Chappuis ozone bands is about 0.17 atm, the corresponding altitude $h_0$ is 12 km. It allows us to compare the obtained $O_3$ and $NO_2$ molecule numbers with SCIAMACHY limb measurements [5, 23, 24] for the same date, integrating the data along the trajectory. The results are shown by the dashed lines in the Figure 4. According to our observations during the eclipse, the abundance of ozone is higher than the same abundance from SCIAMACHY data. It can be explained by the accuracy of estimation of both our measurements (especially for $NO_2$) and SCIAMACHY data, where the relative error of $[O_3]$ and $[NO_2]$ near 10 km is about several dozens of percents.

**5. Water vapor measurements.**

Observed spectral range covers the $H_2O$ absorption bands around 730 nm. This band consists of a number of strong separated lines and can be used to measure the tangent column density of $H_2O$ along the emission path. The absorption of water vapor takes place at the both atmospheric fractions of the path – southern polar stratosphere and troposphere above the observatory.

Figure 8 shows the fragment of the lunar surface spectra near 730 nm inside and outside the umbra. In eclipse-free case just the local troposphere contributes to the absorption. The $H_2O$ lines are naturally weaker than during the eclipse, but the analysis shows the similar line profile characteristics, typical for higher atmosphere levels at lower pressure. The possible reason for that are current special properties of water vapor distribution above the observatory during the observations.

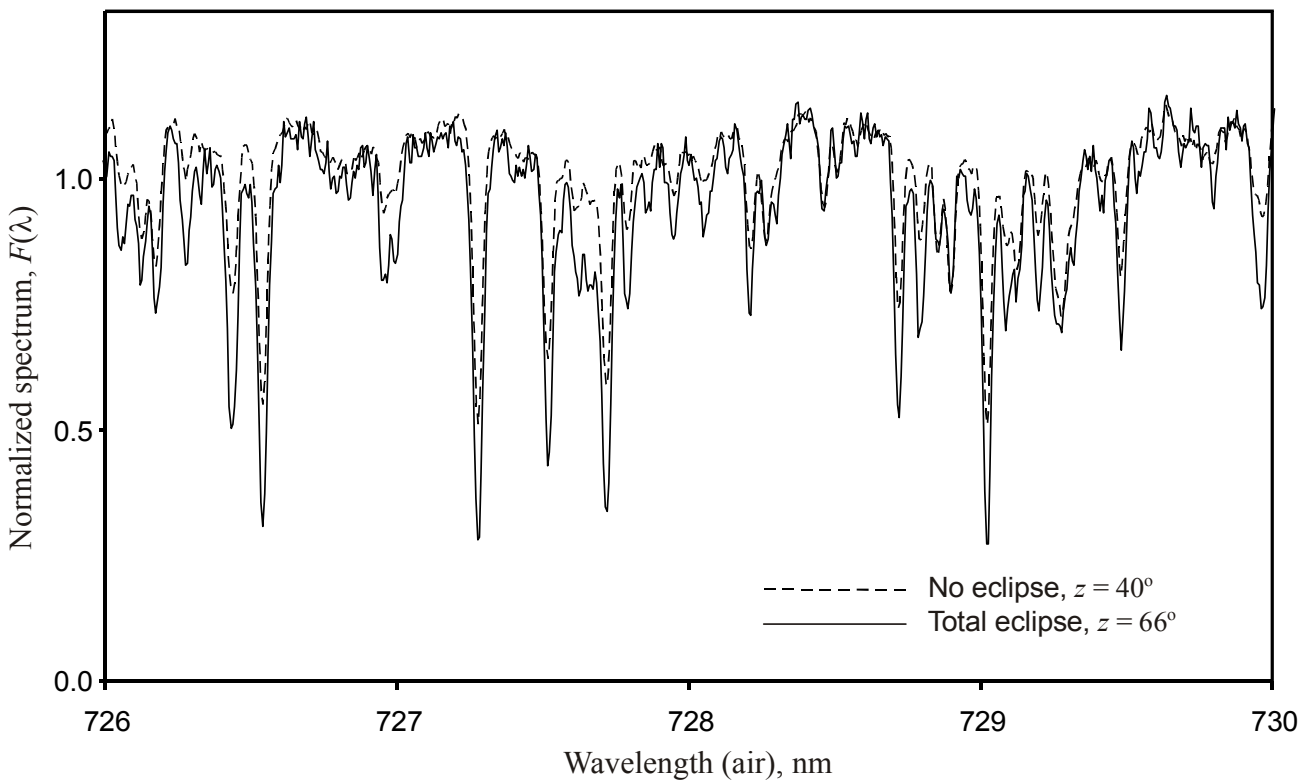

*Figure 8. The fragment of lunar surface spectra inside (spectrum 4) and outside the umbra with $H_2O$ bands near 730 nm.*

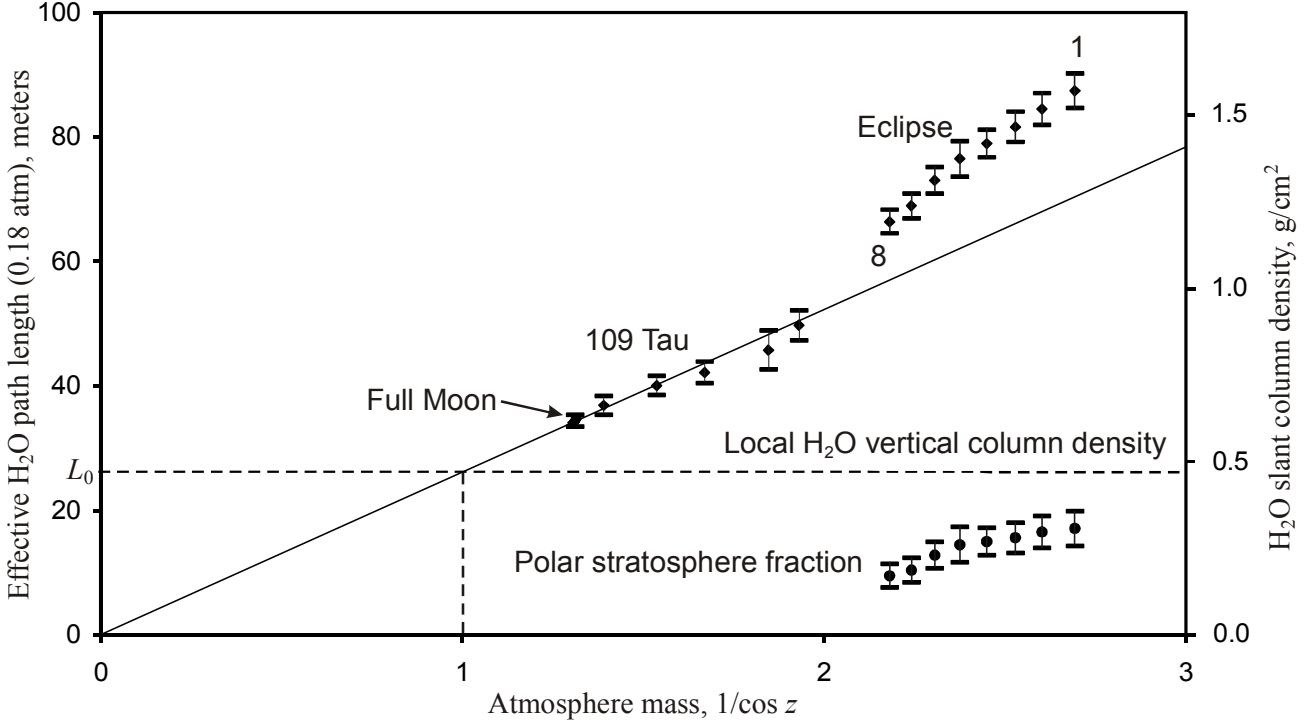

*Figure 9. The abundance of $H_2O$ in the polar stratosphere and local atmosphere based on the observations.*

This circumstance allows us to simplify the numerical analysis. We consider the line profile to be the same along the whole trajectory. Since we are interested mainly in the polar stratosphere and have to separate corresponding $H_2O$ absorption, we take the HITRAN profile [20, 21] for the temperature $T_0$ (220 K) and the effective pressure $p_0$ obtained in previous section for this spectral range, which is 0.18 atm. The spectrum corresponding to the $H_2O$ effective path length $L(z)$ is convolved with the instrumental profile:

$$F_{EL}(\lambda) = \int_{-\lambda_{PSF}}^{\lambda_{PSF}} S(\lambda+\lambda_i)L(\lambda+\lambda_i)E(\lambda)A(\lambda,z)\exp(-X_{H2O}(p_0,T_0,\lambda+\lambda_i)\cdot L(z))\,Q(\lambda,\lambda_i)d\lambda_i \tag{10}$$

Here $E(\lambda)$ and $A(\lambda, z)$ are the continuum light extinction by the paths 1 and 2 (that may be considered as polynomial), $X_{H2O}$ is the water cross-section integrated using the HITRAN database. $L(z)$ is to be found by the least squares method by comparison with observational data. To exclude the local troposphere contribution, we have to run this procedure not only for the umbral spectra, but also for full Moon and standard star ones, covering the wide range of zenith angles of the source.

The results can be seen in the bottom of Figure 9. Observations of uneclipsed Moon and standard star cover the range of atmospheric masses from 1.3 to 2.0. For all these spectra the effective $H_2O$ path length (or corresponding slant column density) is proportional to the atmospheric mass (or $1/\cos z$). It shows the stable local condition of water vapor during the observations. The value of vertical $H_2O$ column density above the observatory is found to be 0.47 g/cm$^2$, corresponding to the length $L_0$ 26.1 meters for the pressure and temperature noted above. The local $H_2O$ vertical column density slightly exceeds to the SCIAMACHY AMC-DOAS data value for this region ([25], 0.34 g/cm$^2$). The effective length of the stratosphere water path can be found using the Bouger's law:

$$L_S = L(z) - \frac{L_0}{\cos z} \tag{11}$$

Dependency of this value is also shown in Figure 9. Typical $L_S$ amount for most part of recorded spectra is 15 meters that corresponds to the tangent column density 0.27 g/cm$^2$ for the pressure 0.18 atm and temperature 220 K. The $H_2O$ molecule number slightly increases in the western part of the limb, where the ozone abundance along the same trajectory decreases.

Given the $H_2O$ vertical column density by SCIAMACHY AMC-DOAS data [25] for the most part of the limb locations (about 0.6 g/cm$^2$), we estimated the characteristic scale height of $H_2O$ distribution by the method described in [7]. The obtained amount is about 2.0 km. It is higher than the tropical latitudes value by the eclipse of March, 4, 2007, but the stratosphere $H_2O$ abundance and the scale value obtained in that photometric paper (1.3 km, [7]) seem to be underestimated due to the Forbes effect (underestimation of fine structure absorption by wide spectral band photometry) caused by the structure of $H_2O$ bands. High-resolution spectroscopy allows us to improve the accuracy of the scale value. If the $H_2O$ distribution followed the exponential law with the same scale both in troposphere and stratosphere, then relative $H_2O$ abundance in the southern lower stratosphere would be found to be about 50 ppm. However, as we saw for the observatory location, the $H_2O$ distribution with altitude does not follow this exponential law, and this amount is just the rough estimation that overestimates the value above the tropopause.

**6. Conclusions and perspectives.**

In this paper the analysis of high resolution spectra of the Moon during the lunar eclipse is performed. The scheme of radiation transfer provides the opportunity to study the absorption of the trace gases along the tangent trajectory numerically far from the place where observations were carried out. Usually, this can be done only by space observations. The position of effective trajectory versus the wavelength allows us to estimate the contribution of scattering emission. It is found to be small at the wavelengths above 500 nm. At this spectral range the eclipsed Moon is quite bright and the large number of atmospheric absorption bands appears.

The lunar eclipse spectra were obtained for the wavelengths up to 780 nm, including $O_2$ and $H_2O$ bands. However, if it is expanded to the IR-range, the lines of greenhouse gases $CO_2$ and $CH_4$ will be included to the analysis [2]. By increasing the number of points of lunar surface or by using the spectrograph with the slit oriented along the umbra radius, we can analyze different trajectories simultaneously and study the vertical distribution of optically active gases.

**Acknowledgements**

Authors would like to thank Artem Burdanov, Alexander Popov and Konstantin Mironov (Kourovka Astronomical Observatory) for their help in preparation and conducting of observations. We also thank Christian von Savigny and Stefan Noël (Institute of Environmental Physics/Remote Sensing, University of Bremen, Germany) for providing the $O_3$, $NO_2$ and $H_2O$ SCIAMACHY data, Boris Voronin, Semen Mikhailenko (Institute of Atmosphere Optics, Tomsk, Russia) and Oleg Postylyakov (Institute of Atmospheric Physics, Moscow, Russia) for the help on working with SPECTRA system and HITRAN database.

The work was performed with partial support by Russian Federal program "Investigations and Elaborations on Priority courses of Russian Scientific and Technological Complex Development 2007-2012" (State contract №16.518.11.7074) and Russian Foundation for Basic Research (grant №12-05-00501).